\documentclass{article}
\usepackage{amsmath}
\usepackage{bm}

\newcommand{\NEG}[1]{\not{#1}}

\begin{document}

\title{Quantum Corrections to Lorentz Invariance Violating Theories:
Fine-Tuning Problem}
\author{P. M. Crichigno and H. Vucetich}
\maketitle

\begin{abstract}
It is of general agreement that a quantum gravity theory will most probably
mean a breakdown of the standard structure of space-time at the Planck
scale. This has motivated the study of Planck-scale Lorentz Invariance
Violating (LIV) theories and the search for its observational signals. Yet,
it has been recently shown that, in a simple scalar-spinor Yukawa theory,
radiative corrections to tree-level Planck-scale LIV theories can induce
large Lorentz violations at low energies, in strong contradiction with
experiment, unless an unnatural fine-tuning mechanism is present. In this
letter, we show the calculation of the electron self-energy in the framework
given by the Myers-Pospelov model for a Lorentz Invariance Violating QED. We
find a contribution that depends on the prefered's frame four-velocity which
is not Planck-scale suppressed, showing that this model suffers from the
same \textit{disease}. Comparison with Hughes-Drever experiments requires a
fine-tuning of 21 orders of magnitude for this model not to disagree with
experiment.
\end{abstract}

\section{Introduction}

\bigskip The two astonishingly successful pillars of contemporary physics,
namely Quantum Mechanics and General Relativity have stubbornly resisted
their unification. The problem of finding a consistent theory that merges
these two aspects of Nature, i.e. a quantum theory of gravitation, is known
as the Quantum Gravity (QG) problem. Currently, the two most important
approaches to this challenge are String Theory, which also claims to be a
complete unification of all forces, and the less ambitious\ theory, known as
Loop Quantum Gravity which is a canonical quantum version of General
Relativity. Although based on different approaches and hypotheses, these
theories agree in that novel properties of the structure and symmetry of
space-time at the Planck-scale must arise, making the experimental
exploration of Planck-scale physics of fundamental importance. Quantum
Gravity phenomenology had been left aside for a while since directly probing
Planck-scale effects on elementary particles would require energies of the
order of $E_{Pl}=\sqrt{\hbar c^{5}/G}\simeq 10^{19}GeV$, which is currently,
and will most probably be for a long time, inaccessible to experimentalists.
Nevertheless, Quantum Gravity phenomenology has enjoyed a renewed interest
in recent years and has actually become a very active research area by the
hand of Amelino Camelia, among others, when it was noticed that some
Planck-scale effects may be amplified in several experimental and
astrophysical scenarios to the extent of becoming observable with current
experimental technology \cite{Amelino Camelia 1},\cite{Amelino Camelia 2}.

Lorentz invariance breakdown at the Planck scale, manifested as a modified
dispersion relation, is one of the common features that arises in the two
mentioned contenders for QG\ phenomenology (see, for instance, \cite{Gambini
Pullin}, \cite{Ellis} ) and one which has attracted considerable attention
lately. Lorentz invariance violation (LIV) signifies a modification of the
group structure of space-time symmetry and so the usual algebra Casimir $%
p_{\mu }\eta ^{\mu \nu }p_{\nu }=m^{2}$ fails to be an invariant under the
set of transformations relating inertial observers in the effective QG
theory, i.e. a modified dispersion relation is expected. In fact, the
existence of a natural mass scale $(M_{Pl}=\sqrt{\hbar c/G})$ allows the
appearance of Planck-scale suppressed terms of powers greater than $2$ in
momenta in the dispersion relation. These terms can be traced back as having
their origin in a modified Lagrangian for fields that contains five(or
higher)-dimensional (i.e. non-renormalizable) kinetic terms. This
field-point of view suggests that one should consider radiative corrections
to the bare vertices of the theory to explore quantum modifications.

The motivation of this letter is the following \cite{Pipi}: The calculation
of loop-diagrams requires the integration over arbitrarily high momenta and
so the high energy regime of the theory is explored. In the usual Lorentz
invariant gauge theories, these diagrams are divergent due to dimension four
(or smaller) terms in the Lagrangian, which are made finite by the
well-known procedure of renormalization. The incorporation of operators of
higher dimension (higher powers in momenta in the dispersion relation)
naturally introduces a mass scale (e.g. the Planck scale) and some of the
originally divergent diagrams are naturally regularized. The superficial
degrees of divergence of the Lorentz invariant diagrams therefore determine
the size of the contributions of the Lorentz violating terms, which involve
the Planck scale. These contributions are not necessarily Planck-scale
suppressed, as frequently stated in the literature. In conclusion, Lorentz
violation at the Planck scale is\textit{\ pulled }down to low energies due
to radiative effects, requiring an enormous fine-tuning in order to be
consistent with experimental bounds on Lorentz violation.

\bigskip

In this letter we show that a modified model of QED, known as the
Myers-Pospelov model, suffers this \textit{disease} and estimate the order
of magnitude of the fine-tuning by comparison with Hughes-Drever
measurements \cite{Exp}, which is in agreement with the general conclusions
of reference \cite{Pipi}. The organization of the paper is as follows:
presentation of the Myers-Pospelov model, calculation of the electron's self
energy diagram and finally the discussion of experimental bounds and the
fine-tuning problem that this arises.

\section{The Myers-Pospelov Model}

\label{sec:MPMod}

The Myers-Pospelov (MP) model, which incorporates cubic modification terms
in the dispersion relation by introducing a background four-vector $w_{\mu
}, $ has received considerable attention lately and exhibits more appealing
properties regarding causality conditions than other LIV\ models \cite%
{Causality}. It has been studied in the language of effective field theory
by introducing dimension-five operators into the lagrangian. More precisely,
the MP's low energy effective field theory introduces Planck's
scale-suppressed operators which satisfy the following general constraints:
Quadratic in the same field, one more derivative than the usual kinetic
term, gauge invariance, Lorentz invariant except for the appearance of $%
w^{\mu }$, not reducible to lower dimension by the equations of motion and
not reducible to a total derivative. The MP Lagrangian describing the
electromagnetic interactions reads \cite{Myers-Pospelov}:

\begin{equation}
\mathcal{L}_{QED}^{MP}=\mathcal{L}_{QED}+\delta \mathcal{L}_{\gamma }+\delta 
\mathcal{L}_{\Psi },
\end{equation}%
wherein $\mathcal{L}_{QED}=-\frac{1}{4}F^{2}+\bar{\Psi}(\not{p}-e\not{A}%
-m)\Psi $ is the usual QED Lagrangian and 
\begin{equation}
\delta \mathcal{L}_{\gamma }=\frac{\xi }{M_{Pl}}w^{a}F_{ad}w\cdot \partial
(w_{b}\tilde{F}^{bd}),
\end{equation}%
\begin{equation}
\delta \mathcal{L}_{\Psi }=\frac{i}{M_{Pl}}\bar{\Psi}(\eta _{1}\not{w}+\eta
_{2}\not{w})(n\cdot \partial )^{2}\Psi .
\end{equation}%
The photon dispersion relation due to this modified Lagrangian is

\begin{equation}
\left( E^{2}-|\bm{p}|^{2}\pm \frac{2\xi }{M_{Pl}}|\bm{p}|^{3}\right)
(\varepsilon _{x}\pm i\varepsilon _{y})=0,
\end{equation}%
while 
\begin{equation}
\left( E^{2}-|\bm{p}|^{2}-m^{2}+\frac{1}{M_{Pl}}|\bm{p}|^{3}(\eta _{1}+\eta
_{2}\gamma _{5})\right) \Psi =0
\end{equation}%
is the modified dispersion relation for electrons in the privileged system
of reference, i.e., the one in which $w^{\mu }=(1,0,0,0)$ meaning that
Poincare's symmetry has been broken to $O(3)$ in this frame. Note that in
this frame the modifications to the usual dispersion relation are related to
the space-components of $p$ while the energy dependence is the usual one.
From these expressions, we can perform the calculation of the electron
self-energy diagram in which we are interested now.

\bigskip

\section{Electron self-energy}

Since it is expected that departures from the (incredibly accurate) usual
QED\ theory manifest themselves at high energies, due to it being a quantum
gravity effect, we shall consider QED as a good enough approximation for 
\textit{energies} smaller than a given energy scale which we shall call $%
\Lambda $, while being the MP theory the accurate description for high
energies compared to $\Lambda $. Then, we shall express the quantum
correction to a given diagram by the symbolical expression%
\begin{equation}
\Sigma =\int_{0}^{\Lambda }d\Sigma _{QED}+\int_{\Lambda }^{\infty }d\Sigma
_{MP}.
\end{equation}%
That is to say that we may think of $\delta \Sigma (\xi ,\Lambda )\equiv
\int_{\Lambda }^{\infty }d\Sigma _{MP}$ as being a \textit{correction} term
to the (regularized) self-energy diagram in QED due to quantum gravity
effects of the MP model. We wish to stress the fact that the $\Lambda $
scale introduced here has a two-fold function: It regularizes the usual
divergent integral of QED and it allows the calculation of $\delta \Sigma
(\xi ,\Lambda )$ in the high-energy regime\footnote{%
More precisely, the $\Lambda $ scale marks the begining of the regime in
which the quantum gravity effects are important. In the privileged reference
frame this occurs for big values of $|\bm{p}|.$ Also, observe that $\Lambda $
acts as an infrared regulator.}. This \textit{cut-off} is of a physical
character and unnecessary at the fundamental level, but it enormously
simplifies the calculations and it allows us to identify more clearly the
modification the MP model introduces. As we shall see, this correction term
is not suppressed by any powers of $1/M_{Pl}$, as could of been naively
expected, giving the finite contribution to the Lorentz violation at low
energies to which we have referred above.

Let us consider the case in which the departure from the usual dispersion
relation is mostly dominated by the $\xi $ parameter in the photon
dispersion relation, while keeping the electron propagator unmodified (i.e. $%
\eta _{1}=\eta _{2}=0$). We wish to explore the effect of 1-loop corrections
in this scenario, in particular the electron self-energy to first order in $%
\hbar $ given by 
\begin{equation}
\begin{split}
-i\Sigma _{MP}(p)=& (-ie)^{2}\frac{1}{(2\pi )^{4}}\int d^{3}k\int
dk_{0}\gamma ^{\mu }\frac{(\not{p}-\not{k}+m)}{(p_{0}-k_{0})^{2}-[(\bm{p}-%
\bm{k})^{2}+m^{2}]}\gamma _{\mu } \\
& \frac{1}{k_{0}^{2}-[\bm{k}^{2}\mp \frac{2\xi }{M_{Pl}}|\bm{k}|^{3}+\lambda
^{2}]}.
\end{split}
\label{self-energy}
\end{equation}

We recall that the superficial degree of divergence of the electron
self-energy diagram in QED\ is 1 and so a linear divergence is to be
expected, but because of Ward's identities this divergence is smoothed into
a logarithmic one to match that of the interaction vertex diagram. Since
Ward's identities are a consequence of gauge invariance, the same can be
said here and it's required that this integral be convergent, as the
interaction vertex is finite in the MP model. Indeed, note that by
regularizing the integral in such a way that the symmetry $k^{\mu
}\rightarrow -k^{\mu }$ at high energies is preserved and writing
the integrand as $I(k)=\frac{1}{2}\left[ I(k)+I(-k)\right] $, one
gets a finite result. That is, the MP model predicts a finite correction to
the self-energy diagram. After using the standard Feynman parameterization
and performing some integrals, we get (see Appendix)%
\begin{equation}
-i\Sigma _{MP}(p)\simeq -i\frac{e^{2}}{8\pi ^{2}}\left( -\NEG{p}-\gamma
^{i}p_{i}+4m\right) \ln \frac{\xi \Lambda }{2M_{PL}}.  \label{correction}
\end{equation}%
Notice that the contribution 
\begin{equation}
\frac{e^{2}}{8\pi ^{2}}\left( -\NEG{p}+4m\right) \ln \frac{\xi \Lambda }{%
2M_{PL}}
\end{equation}%
can be absorbed by the regularized self-energy term coming from QED, leaving 
\begin{equation}
\delta\Sigma_{MP} = \frac{e^{2}}{8\pi ^{2}}\gamma ^{i}p_{i}\ln
\frac{\xi \Lambda }{2M_{PL}} 
\label{finite:corr}
\end{equation}%
as a new and finite term, due to high energy corrections in the MP
model.

Note that (\ref{finite:corr}) has the asymptotic form $f(m/M_P)\ln\xi$
for small $\xi$ and the form $g(\xi) \ln(m/M_P)$ for small
$m/M_P$. Since the latter quantity is $m/M_P \sim 10^{-22}$ we shall
have the behaviour $\delta\Sigma_{MP} \propto g(\xi)
\ln(m/M_P)$. Thus, the very small quantity $m/M_P$ appears as the
argument of a logarithm.

Note
also that the $\Lambda$ dependence on equation (\ref{finite:corr})
appears because a cutoff in momentum space does not preserve the
symmetry $k\to-k$.

Now, recall that this expression is to be understood in the preferred system
of reference in which $w_{\mu }=(1,0,0,0)$, but in an arbitrary reference
frame with relative velocity $w^{\mu }$, we must replace $p^{\iota }$ by the
projection of $p^{\mu }$ onto the transversal direction of $w_{\mu }$, i.e., 
$p_{\mu }^{T}=\Pi _{\mu }^{\nu }(w)p_{\nu },$ where $\Pi _{\mu }^{\nu
}(w)\equiv \delta _{\mu }^{\nu }-w_{\mu }w^{\nu }$ is the usual transversal
projector, if the $w$ four-vector is normalized and $g_{\mu \nu
}=diag(+,-,-,-)$. Finally 
\begin{equation}
\delta \Sigma _{MP}(p,w)=\frac{e^{2}}{8\pi ^{2}}\gamma ^{\mu }(g_{\mu \nu
}-w_{\mu }w_{\nu })p^{\nu }\ln \frac{\xi \Lambda }{2M_{Pl}},
\end{equation}%
wherein $w^{\mu }$ represents the 4-velocity of the observer relative to the
preferred system of reference. Note that the limit $\xi \rightarrow 0$ must
be taken simultaneously with the limit $\Lambda \rightarrow \infty $ in such
a way that $\xi \Lambda $ remains finite in order to recover the low-energy
limit. Notice that the parameter $\xi $, which controls the
extent of Lorentz Invariance violation, regularizes the usual logarithmic
divergence, as could of been expected from power counting. This slightly
resembles what happens in Non-Commutative Field Theory, where the
non-commutativity parameter $\theta $ measures the extent of
Lorentz invariance violation and regularizes some ultravioletly-divergent
diagrams as well \cite{Grandi},\cite{NC FT}.

\section{Lagrangian Counterterm: fine-tuning problem}

If the MP\ model with 1-loop corrections is to be consistent with
experimental bounds on low energy Lorentz violation, it is necessary to
include a (Lorentz violating) counterterm in the original Lagrangian in the
way%
\begin{equation}
\mathcal{L}_{MP}^{\prime }=\mathcal{L}_{MP}^{\mathrm{1-Loop}}-\frac{ie^{2}}{%
8\pi ^{2}}\ln (\frac{\xi ^{\prime }\Lambda }{M_{Pl}})\bar{\Psi}\left( \gamma
^{\mu }(g_{\mu \nu }-w_{\mu }w_{\nu })\partial ^{\nu }\right) \Psi .
\end{equation}%
By introducing $\delta \equiv \frac{\xi }{\xi ^{\prime }}-1$, we have%
\begin{eqnarray}
\mathcal{L}_{MP}^{\prime } &=&\mathcal{L}_{MP}^{0}-\frac{ie^{2}}{8\pi ^{2}}%
\ln (\delta +1)\bar{\Psi}\left( \gamma ^{\mu }(g_{\mu \nu }-w_{\mu }w_{\nu
})\partial ^{\nu }\right) \Psi \\
\mathcal{L}_{MP}^{\prime } &\simeq &\mathcal{L}_{MP}^{0}-\frac{i\delta e^{2}%
}{8\pi ^{2}}\bar{\Psi}\left( \gamma ^{\mu }(g_{\mu \nu }-w_{\mu }w_{\nu
})\partial ^{\nu }\right) \Psi .
\end{eqnarray}%
Notice that the dependence on $\Lambda $ has disappeared and there is no
Planck-scale suppression, as advertised before. The $\delta $ parameter
therefore measures the extent of Lorentz violation, rising the problem of a
fine-tuning of the counterparameter $\xi ^{\prime }$.

\section{Experimental bounds}

The associated Hamiltonian in the non-relativistic limit of this Lagrangian,
given by the method developed in \cite{Kostelecky} up to first order in $%
\delta $, reads%
\begin{equation}
\begin{split}
H= \left[ mc^{2}\left(1-\frac{\delta e^{2}}{4\pi ^{2}}(\mathbf{w}%
/c)^{2}\right)\right. \\
\left. +\left(1-2\frac{\delta e^{2}}{4\pi ^{2}}\left( 1+\frac{5}{6}(\mathbf{w%
}/c)^{2}\right) \right) \left( \frac{p^{2}}{2m}+g\mu \mathbf{s}\cdot \mathbf{%
B}\right) \right] \\
-\frac{\delta e^{2}}{4\pi ^{2}}\left[ \frac{\mathbf{w\cdot }Q_{P}\cdot 
\mathbf{w}}{mc^{2}}\right] .
\end{split}%
\end{equation}

The last term badly breaks Lorentz invariance. It represents an anisotropy
of inertial mass and it has been tested in Hughes-Drever like experiments 
\cite{Exp}. On account of the approximation $Q_{P}=-5/3<p^{2}>Q/R^{2}$ for
the momentum quadrupole moment, with $Q$ being the electric quadrupole
moment, we get%
\begin{equation}
\delta H_{Q}=-\frac{\delta e^{2}}{4\pi ^{2}}\frac{5}{3}\left\langle \frac{%
p^{2}}{2M}\right\rangle \left( \frac{Q}{R^{2}}\right) \left( \frac{w}{c}%
\right) ^{2}P_{2}(\cos \theta ).
\end{equation}%
for the perturbing Hamiltonian, from which we get a bound for the fine
tuning parameter $\delta $ 
\begin{equation}
\left\vert \delta \right\vert <10^{-21}.  \label{eq:delta:Bound}
\end{equation}

\section{Conclusions}

We have explicitly shown the realization, in the framework of the
Myers-Pospelov model for a LIV QED of the idea, first exposed in
\cite{Pipi}, that LIV at the Planck scale at the classical
$(i.e.\,\hbar =0)$ level are dragged down to low energies by radiative
corrections. This is reflected by the fact, already noted, that the
dependence of the LIV term (\ref{finite:corr}) on the very small
factor $m/M_P$ is only logarithmic.  Thus, the LIV is actually
amplified in such a way that a LIV finite counterterm has to be added
to suppress the large (order of the QED fine structure constant) term
and an extreme fine-tuning is required in order for this theory to be
consistent with present experimental bounds.

By comparison with Hughes-Drever measurements\cite{Exp}, we have been
able to determine the magnitude of the fine-tuning to be
\eqref{eq:delta:Bound}, in agreement with the general conclusions of
reference \cite{Pipi}.

Fine-tuning problems already arise within the Standard
model, considered one of the most succesfull theories, but it is expected to
be solved at a more fundamental level and has been considered unacceptable
in a fundamental theory by some authors \cite{Fine Tuning Weinberg},\cite%
{Fine tuning Susskind}. The fine-tuning problem discussed here could also be
considered as a sign suggesting, a still unknown, underlying structure.The
implications of this study are quite restrictive on possible Lorentz
violating scenarios; if the exact Lorentz invariance hypothesis is relaxed
at the Planck-scale, then a deeper and precise mechanism for ensuring
Lorentz invariance at low energies when quantum corrections are considered,
is needed. Some mechanisms to deal with this problem have been proposed, but
none of these are, at present, fully satisfactory. The reader is referred to 
\cite{Collins} for an updated review and discussion of these proposals and,
in general, of LIV's role in quantum gravity phenomenology.

\textbf{\bigskip }

\textbf{ACKNOWLEDGMENTS}

\bigskip The authors would like to thank the (anonymous) reviewer and N.
Grandi for very useful and enlightening comments.

\begin{center}
\textbf{APPENDIX}
\end{center}

We show in this appendix the calculations leading to (\ref{correction}). The
Feynman parameterization 
\begin{equation}
\frac{1}{ab}=\int_{0}^{1}dz\frac{1}{\left( az+b[1-z]\right) ^{2}}
\end{equation}%
allows us to express the integral in (\ref{self-energy}) as 
\begin{equation}
I=\int dz\int dk_{0}\frac{-2\not{p}+4m+2\not{k}}{\left( \left[
p^{2}-2pk-m^{2}\right] z+k^{2}+\left[ -\lambda ^{2}+\frac{2\xi }{M_{Pl}}|%
\bm{k}|^{3}\right] (1-z)\right) ^{2}},
\end{equation}%
where the standard Dirac gamma matrices properties $\gamma ^{\mu }\not{p}%
\gamma _{\mu }=-2\not{p}$ and $\gamma ^{\mu }\gamma _{\mu }=4$ have been
used. By shifting solely the time-like integration variable $k_{0}$ to $%
k^{0}-p^{0}z$, we get%
\begin{equation}
I=\int dz\int dk_{0}\frac{-2\not{p}(1-z)+4m+2\gamma ^{i}(k_{i}+p_{i}z)}{%
\left( k_{0}^{2}+p_{0}^{2}z(1-z)-(\bm{p}-\bm{k})^{2}z-m^{2}z-\left( \bm{k}%
^{2}+\lambda ^{2}-\frac{2\xi }{M_{Pl}}|\bm{k}|^{3}\right) (1-z)\right) ^{2}},
\end{equation}%
where we have used the fact that the integral with $\gamma ^{0}k_{0}$ is
exactly zero, due to the integrand being parity-odd in $k_{0}$. The table
expression 
\begin{equation}
I=\int \frac{d^{d}k}{\left( k^{2}+2kq-r^{2}\right) ^{\alpha }}%
=(-1)^{d/2}i\pi ^{d/2}\frac{\Gamma (\alpha -\frac{d}{2})}{\Gamma (\alpha )}%
\frac{1}{[-q^{2}-r^{2}]^{\alpha -d/2}}
\end{equation}%
allows the full computation of the integral in QED. Although not so useful
now, it allows the whole integration of the time-like component of $k$, by
taking 
\begin{eqnarray}
d &=&1,\quad \alpha =2, \\
r^{2} &=&-p_{0}^{2}z(1-z)+(\bm{p}-\bm{k})^{2}z+m^{2}z+\left( \bm{k}%
^{2}+\lambda ^{2}-\frac{2\xi }{M_{Pl}}|\bm{k}|^{3}\right) (1-z) \\
q &=&0.
\end{eqnarray}%
This way, the integral to be performed over the spatial components of $k$ is

\begin{equation}
-\frac{\pi }{2}\int d^{3}k\frac{-2\not{p}(1-z)+4m+2\gamma ^{i}(k_{i}+p_{i}z)%
}{\left[ p_{0}^{2}z(1-z)-(\bm{p}-\bm{k})^{2}z-m^{2}z+\left( -\bm{k}%
^{2}-\lambda ^{2}+\frac{2\xi }{M_{Pl}}|\bm{k}|^{3}\right) (1-z)\right] ^{3/2}%
}.  \label{sin k}
\end{equation}%
It is important to stress that no shifting in the spatial components of $k$
has been done, \ but solely in the time component, $k_{0}$, which is
consistent with the lagrangian's symmetry and the integral being calculated
as the energy integral has no intermediate cut-off.

\bigskip

By taking the external electron to be on-shell we have

\begin{multline}
I=-\frac{\pi }{2}\left( \int d^{3}k\frac{-2\NEG{p}(1-z)+4m-2\gamma ^{i}p_{i}z%
}{\left[ -p_{0}^{2}z^{2}-k^{2}+2pkz\cos \theta +\frac{2\xi }{M_{Pl}}%
(1-z)k^{3}\right] ^{3/2}}\right.  \\
\left. +\int d^{3}k\frac{2\gamma ^{i}k_{i}}{\left[ p_{0}^{2}z(1-z)-(\bm{p}-%
\bm{k})^{2}z-m^{2}z+\left( -\bm{k}^{2}-\lambda ^{2}+\frac{2\xi }{M_{Pl}}|%
\bm{k}|^{3}\right) (1-z)\right] ^{3/2}}\right) ,
\end{multline}%
\begin{equation}
\equiv -\frac{\pi }{2}\left( I_{Log.}+I_{Lin.}\right) ,
\end{equation}%
where the infrared regularization parameter $\lambda $ has been omitted
because this expression is valid at high energies. \ We have already argued
that the $I_{Lin.}$ $=0$ sow we perform the calculation of the (originally)
logarithmically divergent integral, $I_{Log}$. We may first perform the
angular integral over $d\Omega =-d(\cos \theta )d\phi $, getting%
\begin{multline}
I_{Log.}=-2\pi \left( -2\NEG{p}(1-z)+4m-2\gamma ^{i}p_{i}z\right) \cdot  \\
\int dkk\frac{2}{2|\bm{p}|z}\left( \frac{1}{\sqrt{\frac{2\xi }{M_{Pl}}%
(1-z)k^{3}-k^{2}+2pkz-p_{0}^{2}z^{2}}}\right.  \\
\left. -\frac{1}{\sqrt{\frac{2\xi }{M_{Pl}}%
(1-z)k^{3}-k^{2}-2pkz-p_{0}^{2}z^{2}}}\right) .
\end{multline}%
By introducing $\varepsilon \equiv 2pkz-p_{0}^{2}z^{2}$ in the first term
and $\varepsilon ^{\prime }=2pkz+p_{0}^{2}z^{2}$ in the second one we have,
to first order in $\varepsilon $ and $\varepsilon ^{\prime }$, 
\begin{eqnarray}
I_{Log} &\simeq &\frac{\pi }{|\bm{p}|z}\left( -2\NEG{p}(1-z)+4m-2\gamma
^{i}p_{i}z\right) \int dk\frac{(\varepsilon +\varepsilon ^{\prime })k}{%
\left( -k^{2}+2k^{3}\frac{\xi }{M_{Pl}}\left( -z+1\right) \right) ^{\frac{3}{%
2}}}  \notag \\
&\simeq &4\pi \left( -2\NEG{p}(1-z)+4m-2\gamma ^{i}p_{i}z\right) \int dk%
\frac{1}{k\left[ \frac{2\xi }{M_{PL}}(1-z)k-1\right] ^{3/2}}.
\end{eqnarray}%
The definite integral between the values $\Lambda $ and $\infty $ , in which
we are interested in, reads%
\begin{eqnarray}
I_{Log} &\simeq &-8\pi \left( -2\NEG{p}(1-z)+4m-2\gamma ^{i}p_{i}z\right)
\cdot   \notag \\
&&\left( \frac{\pi }{2}-\frac{1}{\sqrt{\frac{2\xi }{M_{PL}}(1-z)\Lambda -1}}%
-\arctan \sqrt{\frac{2\xi }{M_{PL}}(1-z)\Lambda -1}\right)   \notag \\
I_{Log} &\simeq &-4\pi \left( -2\NEG{p}(1-z)+4m-2\gamma ^{i}p_{i}z\right)
\left( \pi +i\ln \frac{\xi \Lambda (1-z)}{2M_{PL}}\right) ,
\end{eqnarray}%
where the standard $\tan ^{-1}(ix)=i\tanh ^{-1}(x)$ and $\tanh ^{-1}(x)=%
\frac{1}{2}\ln (\frac{1+x}{1-x})$ identities have been used in the last line
and the small terms when $\xi \Lambda /M_{PL}\ll 1$ have been neglected.
Integrating over the Feynman parameter $z,$ 
\begin{equation}
\int_{0}^{1}I_{Log}dz=-4\pi i\left( -\NEG{p}-\gamma ^{i}p_{i}+4m\right) \ln 
\frac{\xi \Lambda }{2M_{PL}}+\text{finite terms},
\end{equation}%
which finally gives 
\begin{equation}
-i\Sigma _{MP}(p)=-i\frac{e^{2}}{8\pi ^{2}}\left( -\NEG{p}-\gamma
^{i}p_{i}+4m\right) \ln \frac{\xi \Lambda }{2M_{PL}}.
\end{equation}


\begin{thebibliography}{99}
\bibitem{Amelino Camelia 1} Amelino-Camelia, G., Ellis., J. R., Mavromatos,
N. E., Nanopoulos, D. V., and Sarkar, S., \textit{Nature}, \textbf{393},
763-765 (1998).

\bibitem{Amelino Camelia 2} Amelino-Camelia, G., Introduction to
quantum-gravity phenomenology (2004). ArXiv gr-qc/0412136.

\bibitem{Gambini Pullin} Gambini, R. and Pullin, J. \textit{Phys. Rev., }%
\textbf{D59}, 124021 (1999)

\bibitem{Ellis} Ellis, J. R. Mavromatos, N. E., and Nanopolous, D. V., 
\textit{Gen. Rel. Grav., }\textbf{32}, 127-144 (2000).

\bibitem{Pipi} Collins, J., Perez, A. Sudarsky, D., Urrutia, L. and
Vucetich, H., \textit{Phys. Rev. Lett.}, \textbf{93}, 191301 (2004).

\bibitem{Exp} T.E. Chupp \textit{et al.}, \textit{Phys. Rev. Letts}. \textbf{%
63}, 1541 (1989). V.W. Hughes, H.G. Robinson and V. Beltr\'{a}n-L\'{o}pez, 
\textit{Phys. Rev. Letts}. \textbf{4}, 342 (1960); R.W.P. Drever, Philos.
Mag. \textbf{6}, 683 (1961). J.D. Prestage \textit{et al.}, \textit{Phys.
Rev. Letts}. \textbf{\ 54}, 2387 (1985). S.K. Lamoreaux \textit{et al.}, 
\textit{Phys. Rev. Letts}. \textbf{57}, 3125 (1986); S.K. Lamoreaux \textit{%
et al.}, Phys. Rev. \textbf{\ A 39}, 1082 (1989).

\bibitem{Causality} Martinez, S., Montemayor, R. and \ Urrutia, L. Phys.Rev.
D74 (2006) 065020. ArXiv: gr-qc/0511117.

\bibitem{Myers-Pospelov} R.C Myers and M. Pospelov. ArXiv: gr-qc/0402028.

\bibitem{Grandi} Observation due to N. Grandi in private communications.

\bibitem{NC FT} S. Minwalla, M. Van Raamsdonk, N. Seiberg. ArXiv:
hep-th/9912072 

\bibitem{Kostelecky} V. A. Kostelecky and C. D. Lane, J. Math. Phys. \textbf{%
40}, 6245 (1999).

\bibitem{Supersymmetry} S. Groot Nibbelink and M. Pospelov, Phys. Rev. Lett.
94, 081601 (2005). ArXiv: hep-ph/0404271.

\bibitem{Fine Tuning Weinberg} S. Weinberg, Phys. Rev. D \textbf{8}, 4482
(1973).

\bibitem{Fine tuning Susskind} L. Susskind, Phys. Rev. D \textbf{20}, 2619
(1979).

\bibitem{Sudarsky:2002ue} Sudarsky, D. and Urrutia, L. and Vucetich, H.,%
\textit{\ Phys. Rev. Lett}. \textbf{89} 231301 (2002)

\bibitem{Collins} Collins, J., Perez, A. and Sudarsky, D. Draft chapter
contributed to the book "Towards quantum gravity", being prepared by Daniele
Oriti for Cambridge University Press ArXiv: hep-th/0603002.
\end{thebibliography}
\end{document}